\documentclass[twoside]{report}
\usepackage{iwsm}
\usepackage{graphicx}
\usepackage{amsmath, amssymb}
\usepackage{booktabs}

\begin{document}


\title{Bayesian Generalized Nonlinear Models Offer Basis Free SINDy With Model Uncertainty}
\titlerunning{BGNLM for SINDy}

\author{Aliaksandr Hubin\inst{1}}
\authorrunning{Hubin et al.}    

\institute{NMBU, HiOF, UiO (Integreat), Norway}

\email{aliaksandr.hubin@nmbu.no}

\abstract{Sparse Identification of Nonlinear Dynamics (SINDy) has become a standard methodology for inferring governing equations of dynamical systems from observed data using statistical modeling. However, classical SINDy approaches rely on predefined libraries of candidate functions to model nonlinearities, which limits flexibility and excludes robust uncertainty quantification. This paper proposes Bayesian Generalized Nonlinear Models (BGNLMs) as a principled alternative for more flexible statistical modeling. BGNLMs employ spike-and-slab priors combined with binary inclusion indicators to automatically discover relevant nonlinearities without predefined basis functions. Moreover, BGNLMs quantify uncertainty in selected bases and final model predictions, enabling robust exploration of the model space. In this paper, the BGNLM framework is applied to several three-dimensional (3D) SINDy problems. 
}

\keywords{BGNLM; GMJMCMC; SINDy; Flexible Statistical Modeling.}

\maketitle



\section{Introduction}
The sparse recovery of dynamical systems has recently become a useful tool in applied sciences, with SINDy offering a particularly effective method for reconstructing governing equations (Brunton et al, 2016). However, SINDy simply applies $\ell_1$ regularized lasso variable selection for predefined libraries of candidate basis functions (e.g., polynomials, trigonometric terms), which may limit its adaptability. Furthermore, SINDy requires tuning of the regularization strength for the $\ell_1$ norm and does not quantify uncertainty in the selected models. The latter was resolved in a Bayesian version of SINDy (Hirsh et al, 2022), yet this solution still relies on a predefined library, limiting the flexibility of statistical modeling.  For the former, symbolic regression has recently been applied to SINDy (d'Ascoli et al, 2024), however, it does not handle uncertainty in the selected basis.

Bayesian Generalized Nonlinear Models (BGNLMs) (Hubin et al, 2021) can address both of these challenges. BGNLMs, on one hand, allow for flexible statistical modeling of nonlinear features, but on the other hand, employ a probabilistic framework with spike-and-slab priors, allowing for sparse selection of bases while accounting for uncertainty.  Binary inclusion indicators are used to define the model priors, and Bayesian inference through a genetically modified mode jumping MCMC explores the vast nonlinear model space, providing marginal distributions over plausible governing equations. These capabilities suggest that BGNLM may offer a novel, powerful, and robust tool for solving SINDy problems.

\section{Mathematical Framework}
Consider a system of ordinary differential equations (ODEs) of the form:
\[
\dot{\boldsymbol{x}}(t) = \boldsymbol{f}(\boldsymbol{x}(t)),
\]
where $\boldsymbol{x}(t) \in \mathbb{R}^m$ is the state vector, and $\boldsymbol{f}(\cdot)$ is the unknown governing function. Assume that we observe a sample of noisy realizations of the stochastic process associated with the derivatives $Y_i =\dot{\boldsymbol{x}}(t_i) + \epsilon$ with $Y_i \in \mathbb{R}^m, i \in \{1,\ldots,n\}$ and $\epsilon \sim \mathcal{MVN}(\textbf{O},\sigma^2\mathbf{I_m})$. Further, we observe the process itself at these points ${\boldsymbol{x_i} = \boldsymbol{x}}(t_i),i \in \{1,\ldots,n\}$. 

BGNLM allows to model each component $\dot{{x_j}}(t) = \boldsymbol{f}_j(\boldsymbol{x}(t))$ as $Y_{ji} \sim N(f_j(\boldsymbol{x_i}),\sigma^2), j \in \{1,\ldots,m\}$ with
\[
f_j(\boldsymbol{x_i}) = \sum_{k=1}^{q} \gamma_{jk} \beta_{jk} g_k(\boldsymbol{x_i}),
\]
where
 \( \{g_k(\boldsymbol{x})\}^q_{k=1} \) is the set of BGNLM's nonlinear features, \(\beta_{jk}\in\mathbb{R}\) are coefficients for the \( k \)-th basis of the \( j \)-th equation,
\( \gamma_{jk} \in \{0, 1\} \) are binary inclusion indicators for them. The inclusion indicators \( \gamma_{jk} \) are governed by Bernoulli priors:
$
\gamma_{jk} \sim \text{Bernoulli}(\pi_k),
$
where \( \pi_k \) represents the prior probability of including \( g_k(\boldsymbol{x}) \) in the models, which fully follow Hubin et al (2021). The coefficients \( \beta_{jk} \) are assigned Jeffreys conditional on the configuration of the inclusion of the respective bases  (Hubin et al, 2021), and \( \sigma^2\) parameter is assumed fixed. And \(q\) is the total number of possible features.

Compared to classical SINDy, BGNLM discovers relevant nonlinearities by GMJMCMC and estimates the probabilities of inclusion of the components into the basis. This allows for selecting a robust solution through the median probability model, which selects the features with \( p(\gamma_{jk}=1|\mathcal{D})>0.5\).

\section{Applications}

\subsection{Experimental Systems}
Table~\ref{tab:systems} summarizes three nonlinear systems we will be identifying using statistical modeling with BGNLMs: (a) Linear 3D, (b) Lorenz 3D, and (c) Rössler-Lorenz Hybrid 3D. 

\begin{table}[h!]
\centering
\caption{\small Governing equations, parameters, and initial conditions for 3 systems.}
\label{tab:systems}
\footnotesize
\begin{tabular}{cp{3cm}p{3cm}p{2cm}}
\hline
\textbf{System} & \textbf{Equations} & \textbf{Parameters} & \textbf{Init. Cond.} \\
\hline
Linear 3D & 
\(
\begin{aligned}
\dot{x} &= a x + b xy \\
\dot{y} &= c x +  d xy \\
\dot{z} &= e z
\end{aligned}
\) & 
\(
\begin{aligned}
a &= -1,\textbf{    } b = 20, \\
c &= -20, d = -1,\\
e &= -3
\end{aligned}
\) & 
\(
\begin{aligned}
x &= 2, \\
y &= 0, \\
z &= 1
\end{aligned}
\) \\
\hline
Lorenz 3D & 
\(
\begin{aligned}
\dot{x} &= \sigma(y - x) \\
\dot{y} &= x(\rho - z) - y \\
\dot{z} &= xy - \beta z
\end{aligned}
\) & 
\(
\begin{aligned}
\sigma &= 10, \\
\rho &= 28, \\
\beta &= \frac{8}{3}
\end{aligned}
\) & 
\(
\begin{aligned}
x &= -0.5, \\
y &= -2, \\
z &= 3
\end{aligned}
\) \\
\hline
Hybrid 3D & 
\(
\begin{aligned}
\dot{x} &= -y - z + a \sin(x) \\
\dot{y} &= x + b y \\
\dot{z} &= c + z(x - d)
\end{aligned}
\) & 
\(
\begin{aligned}
a &= 0.2, \\
b &= 0.1, \\
c &= 0.4, d = 5.7\\
\end{aligned}
\) & 
\(
\begin{aligned}
x &= 0.5, \\
y &= -1, \\
z &= 2
\end{aligned}
\) \\
\hline
\end{tabular}
\end{table}


The simulations were performed for the systems using a fine-grained time step of $\Delta t = 0.0001$ and a total simulation time of $T=50$, resulting in $n =500,000$ observations per trajectory. The finite difference method was used to compute derivatives from the trajectories. Further, the simulations were conducted at multiple distinct noise levels, defined as $0.1\times2^k,k\in{0,\ldots,7}$, which were added to the response. Furthermore, 10 independent repetitions were run per noise level, resulting in a comprehensive exploration of model performance under diverse conditions.  

For each replication at each noise level, the dataset was divided into the following subsets for evaluation and training:  \textbf{Training data:} Randomly sampled $1,000$ observations from $t \in (0.05,49.5)$.
\textbf{In-sample predictions:} A subset of 1,000 observations randomly sampled from $t \in (0.05,49.5)$ but non-overlapping with the training data. \textbf{Out-of-sample predictions:} 1000 samples from two disjoint intervals, covering the first and the last parts of the trajectories $t\in[0,0.05]\cup[49.5,50]$, allowing us to estimate the generalizability of the found solutions to completely unobserved parts of the domain. 
For each noise level and repetition, model performance was evaluated based on Power/FDR analysis of identified terms, as well as $R^2$ for the training, in-sample predictions, and out-of-sample prediction sets. The median probability model was used as the final selected model for the identified terms and predictions. Predictions were made using the posterior mode estimates of the parameters of the median probability model under Jeffreys priors with $\sigma^2=1$. The tuning parameters for the \texttt{fbms} function were set as follows (without additional tuning): The population size (\texttt{pop.max}) was limited to 15, with original features retained (\texttt{keep.org = TRUE}). The number of MJMCMC iterations per population was set to 500. Nonlinear transformations included \texttt{sin\_deg}, \texttt{cos\_deg}, and first-order fractional polynomials. The maximal population size was set to 20. Finally, 10 chains of the GMJMCMC were run for every simulation. 
Implementation with all details is available at \texttt
{github.com/aliaksah/BGNLM-for-SINDy}.

\section{Results and Discussion}
\begin{figure}
    \centering
    \includegraphics[trim={0.3cm 1.3cm 0 2cm},clip,width=1\linewidth]{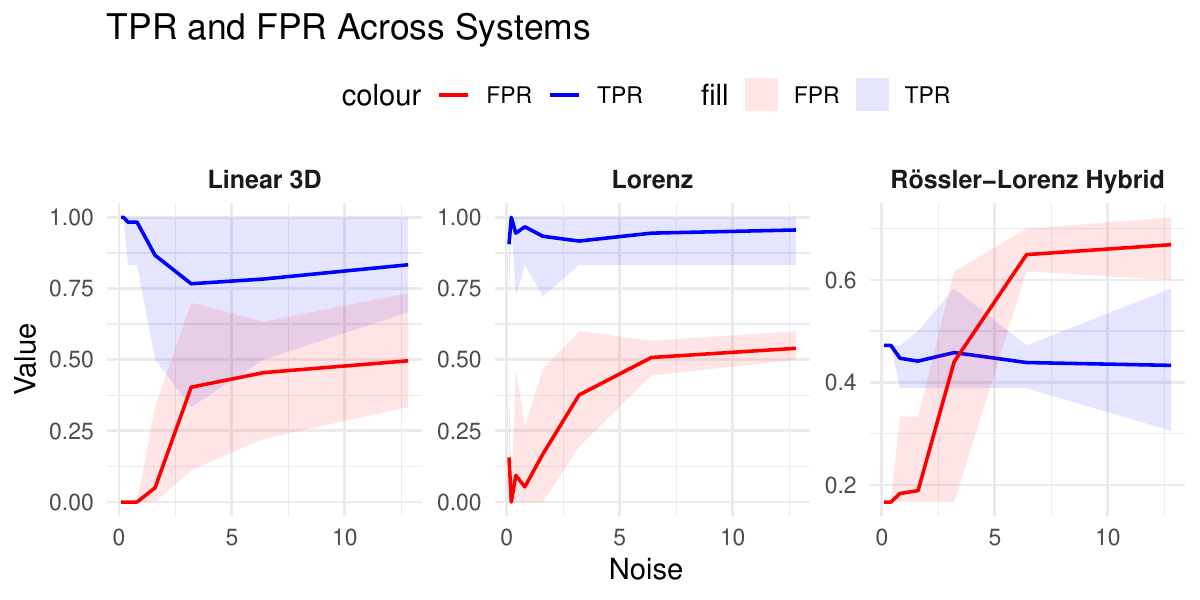}
     \includegraphics[trim={0.15cm 0.2cm 0 3.5cm},clip,width=1\linewidth]{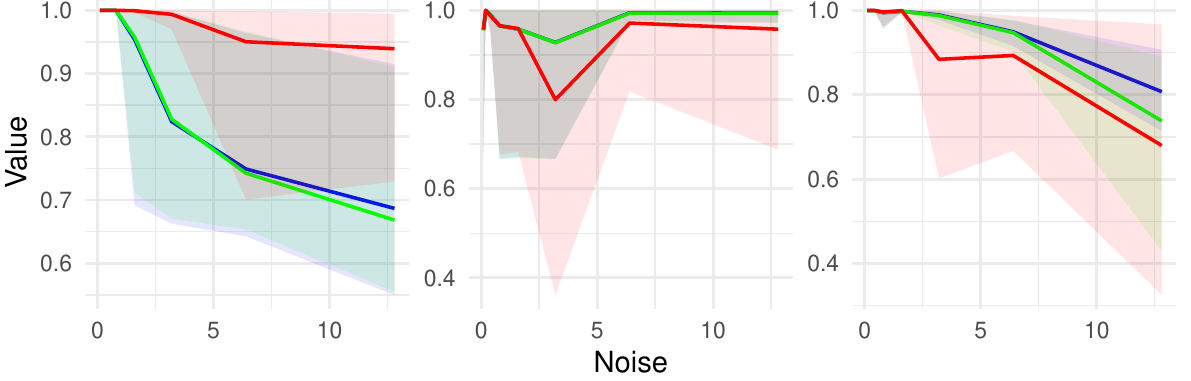}
    \caption{\small \textbf{Top:} Power (blue) FDR (red) curves for the identification of the three systems. \textbf{Bottom:} $R^2$ curves for the predictions of the three systems for train (blue), test (green), and out-of-sample test (red) data.}
    \label{fig:enter-label}
\end{figure}

BGNLM showed a promising trade-off between statistical Power and FDR for identifying the governing equations for each system, especially for lower noise levels, see Figure \ref{fig:enter-label}. Predictions also seem to be reasonable not only for the in-sample domain but also for the out-of-distribution data, yet their quality often drops for higher levels of noise. Thus, flexible Bayesian statistical modeling allows exploration of multiple plausible nonlinear models with evaluated uncertainty. 

In the future, it is of interest to obtain theoretical results for the signal-to-noise ratios allowing for the identifiability of the system. Further, de-noising of the derivatives will become crucial for accurate discoveries under small and sparse samples.  Finally, applications to real-world dynamic systems will be of interest to check the practical use of BGNLMs for applied problems.




\references
\begin{description}
\item[Brunton, S.L., Proctor, J.L., and Kutz, J.N.] (2016).
Discovering governing equations from data: Sparse identification of nonlinear dynamical systems.
\textit{Proc. Natl. Acad. Sci. U.S.A.}, \textbf{113}(15), 3932--3937.

\item[Hirsh, S.M., Barajas-Solano, D.A., and Kutz, J.N.] (2022).
Sparsifying priors for Bayesian uncertainty quantification in model discovery.
\textit{R. Soc. Open Sci.}, \textbf{9}(2), 211823.

\item[Hubin, A., Storvik, G., and Frommlet, F.] (2021).
Flexible Bayesian nonlinear model configuration.
\textit{J. Artif. Intell. Res.}, \textbf{72}, 901--942.

\item[d’Ascoli, S., Becker, S., Schwaller, P., Mathis, A., and Kilbertus, N.] (2024).
ODEFormer: Symbolic Regression of Dynamical Systems with Transformers.
\textit{ICLR 2024}.
\end{description}

\end{document}